\documentclass[11pt]{article}

\usepackage{amssymb,epsfig,fullpage}

\usepackage{subfigure}
\usepackage{graphicx}

\def\arccos{\mathop{\textrm{arccos}}\nolimits}
\def\Arctan{\mathop{\textrm{Arctan}}\nolimits}
\def\Arg{\mathop{\textrm{Arg}}\nolimits}

\begin{document}

\title{Control of unstable steady states in neutral time-delayed systems}

\author{K.B. Blyuss\thanks{Corresponding author. Email: k.blyuss@sussex.ac.uk},\hspace{0.2cm} Y.N. Kyrychko
\\\\ Department of Mathematics, University of Sussex,\\Brighton, BN1 9QH, United Kingdom\\
\and P. H\"ovel and E. Sch\"oll\\\\
Institut f\"ur Theoretische Physik, Technische Universit\"at Berlin,\\ 10623 Berlin, Germany}

\maketitle

\begin{abstract}

We present an analysis of time-delayed feedback control used to stabilize an unstable steady state of a neutral delay differential equation. Stability of the controlled system is addressed by studying the eigenvalue spectrum of a corresponding characteristic equation with two time delays. An analytic expression for the stabilizing control strength is derived in terms of original system parameters and the time delay of the control. Theoretical and numerical results show that the interplay between the
control strength and two time delays provides a number of regions in the parameter space where the time-delayed feedback control can successfully stabilize an otherwise unstable steady state.

\end{abstract}

\section{Introduction}

Control theory plays an important role in engineering, physics and
applied mathematics. It is a well known fact in dynamical systems
theory that any systems can be controlled by applying small
perturbations to the available so-called control parameter. Using a
control theory approach it was shown that it is possible to
control unstable or even chaotic oscillations in many systems
\cite{ogy,hunt,drs}. This possibility of control is of a paramount
importance in many real-life applications, where unwanted
oscillations can lead to poor performance and even dangerous
operating regimes \cite{step01,schlesner}. An excellent overview of recent advances in
the control of nonlinear dynamical systems can be found in Ref.~\cite{book}. One of possible approaches to control unstable steady states is an adaptive stabilization based on a low-pass filter. Details of such a scheme together with examples, including stabilization of a driven pendulum, Lorenz system and an electrochemical oscillator are given in Refs.~\cite{pyr02l,pyr04e}. 
Another control technique, which is quite often used in applications (e.g. optical experiments), and may even be simpler to realize than a low-pass filter, is time-delayed feedback control (TDFC), or time-delay autosynchronization first introduced by
Pyragas in \cite{pyr}; a survey of the developments of this
control method is given in Ref.~\cite{pyr06}. This type of control employs the
difference between the state of the system at
the current time and the state of the system some time $\tau$ ago.
In the problem of stabilizing periodic orbits, the time delay $\tau$
is usually chosen to be the period of the unstable periodic orbit (UPO). This
makes the TDFC non-invasive as it
vanishes as soon as the target state is reached, and it does not
require a lot of information about the system in general. TDFC has been successfully applied to
many physical problems such as, for example, controlling unstable
motion in electronic chaotic oscillators \cite{pyr93,gaut,soc},
control of spatio-temporal UPOs in coupled
reaction-diffusion models \cite{fran,beck02,bert,unkel}. Further
modifications of TDFC include extended
time-delayed feedback control, where multiple time delays
are considered in the feedback loop \cite{soc}, unstable feedback controllers
\cite{pyr01}, and spatio-temporal filtering \cite{baba}.

Besides control of unstable periodic orbits, TDFC and its multiple time delay
extension can be successfully used to control unstable steady states \cite{ahl04,ahl05,hs05,dah07}.
In this case, there is an extra flexibility
in varying the time delay, which in the case of controlling UPOs is required to coincide with a period of an UPO to
ensure that the control is non-invasive. While TDFC is often unable to stabilize a state having
a single real positive eigenvalue \cite{just97} (albeit this is not generally true for UPOs \cite{FIE07}),
it is quite successful in stabilizing steady states with a pair of unstable complex eigenvalues \cite{hs05,dah07}. Asymptotic analysis of
the appropriate characteristic equation shows that, provided the steady state
is close enough to the instability threshold, the
TDFC provides efficient stabilization even for relatively large time delays \cite{yan}. 

In this paper we use time-delayed feedback
control to stabilize an unstable steady state of a neutral delay
differential equation (NDDE). A delay differential equation is
called neutral if it contains a time delay in the highest
derivative involved. This type of equations arises in numerous
physical and engineering application, for example, hybrid testing
\cite{kyr,kyr07}, chaotic oscillations in transmission lines
\cite{bla04c} and torsional waves of a drill string \cite{bal03}.
Due to the original time delay present in the system, its steady states  may become
unstable through a Hopf bifurcation, and adding a time-delayed feedback control will stabilize
the system again. The interplay between the control strength, the
internal and the external time delays, leads to the existence of various regimes in the
parameter space where the steady state of the controlled system is stable. At the same time, due to essential instability in the spectrum, it is impossible to achieve controllability when the value of the coefficient multiplying the highest order time-delayed derivative in the equation exceeds one.

The paper is organized as follows. We first present a stability analysis of an NDDE with a single time delay. Then we introduce time-delayed feedback control to stabilize an unstable steady state of the original system. We show that by varying the control strength and the time delay it is possible to expand the stability region for a range of time-delays. To better understand the influence of TDFC on the stability of the system, we numerically compute the eigenvalue spectrum of the fixed point. The paper concludes with a summary.

\section{Stability analysis}

\begin{figure*}
\hspace{1cm}\includegraphics[width=14cm]{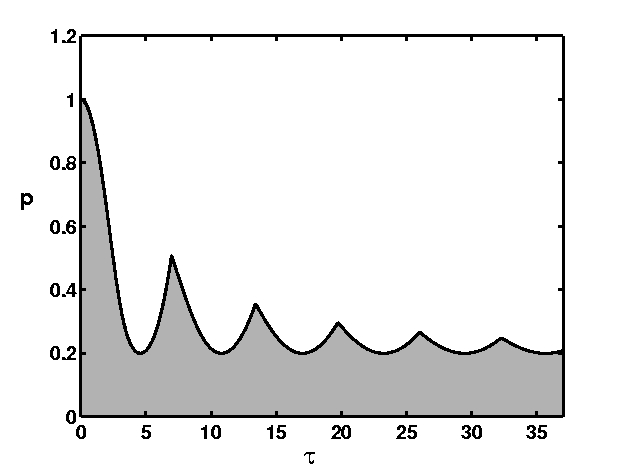}
\caption{Stability boundary of the system (\ref{eq1}) for $\zeta=0.1$. The steady state is stable in the shaded area.}
\label{bound}
\end{figure*}

Consider the following NDDE derived in Kyrychko {\it et al.} \cite{kyr}, which describes a pendulum-mass-spring-damper system:
\begin{equation}\label{eq1}
\ddot{z}(t)+2\zeta\dot{z}(t)+z(t)+p\ddot{z}(t-\tau)=0,
\end{equation}
where dot means differentiation with respect to time $t$, $\tau$ is the time delay, $\zeta$ and $p$ are model parameters corresponding to the damping constant and the ratio of the mass of the pendulum to the mass of the mass-spring-damper, respectively. We will use this equation as a prototype model for the  linearization of other neutral DDEs (see \cite{bla04c} and \cite{bal03} for examples) near their steady states.
This equation has a single steady state $z^{*}=0$, whose
stability is determined by the roots of the corresponding characteristic equation
\begin{equation}\label{char_eq1}
\lambda^{2}+2\zeta\lambda+1+p\lambda^{2}e^{-\lambda\tau}=0.
\end{equation}
In the case when the parameter $p$ in Eq. (\ref{eq1}) exceeds unity, the steady state is unstable for any positive time delay $\tau$. As it was shown in Kyrychko {\it et al.} \cite{kyr}, for $|p|<1$ the steady state switches stability as the time delay is varied. The stability  boundary can be parametrized by the Hopf frequency $\omega$. In this case, the critical time delay and the critical value of $p$ at the stability boundary are given by
\[
\begin{array}{l}
\displaystyle{\tau_{c}=\frac{1}{\omega}\left[\Arctan\frac{2\zeta\omega}{\omega^{2}-1}\pm\pi k\right],}\\\\
\displaystyle{p_{c}=\frac{1}{\omega^{2}}\sqrt{\left(\omega^{2}-1\right)^{2}+4\zeta^{2}\omega^{2}},}
\end{array}
\]
where $k=0$, $1$, $2$, ... and Arctan denotes the principal value of arctan. Figure~\ref{bound} illustrates the dependence of the critical value of $p$ on the time delay $\tau$ which ensures the stability of the steady state. It is noteworthy that if $|p|>1$, the steady state is unstable for any positive time delay $\tau$ due to the so-called essential instability \cite{hale}. On the other hand, if $|p|<1$ and $\zeta>1/\sqrt{2}$, then the steady state is asymptotically stable for any positive time delay $\tau$. For $\zeta<1/\sqrt{2}$, there is a lower bound on the value of $p_{\rm min}=2\zeta\sqrt{1-\zeta^{2}}$, so that asymptotic stability is guaranteed for all $\tau>0$ provided $p<p_{\rm min}$  \cite{kyr}.

In the area above the boundary curve in Fig.~\ref{bound}, the steady state of Eq.~(\ref{eq1}) is unstable, with a pair of complex conjugate eigenvalues in the right complex half-plane. The aim of this study is to stabilize this unstable steady state by means of time-delayed feedback control. To this end, we replace the original system (\ref{eq1}) by its modified version
\begin{equation}\label{eq1_control}
\ddot{z}(t)+2\zeta\dot{z}(t)+z(t)+p\ddot{z}(t-\tau_{1})=K[z(t)-z(t-\tau_{2})],
\end{equation}
where $K>0$ is the strength of the control force, and $\tau_{2}>0$ is the time delay of the control. This modified system has the same steady state as the original system for any time delay $\tau_{2}$, but now the stability of this equilibrium is determined by the roots of the modified characteristic equation:
\begin{equation}\label{char_eq_control}
\lambda^{2}+2\zeta\lambda+1+p\lambda^{2}e^{-\lambda\tau_{1}}=K\left(1-e^{-\lambda\tau_{2}}\right).
\end{equation}
If $K=0$ or $\tau_{2}=0$, we recover the original uncontrolled model with an unstable steady state. The goal is now to find a relation between $K$ and $\tau_{2}$ in terms of the original system parameters $p$, $\zeta$, and $\tau_{1}$, so that in the absence of control there is a pair of unstable complex conjugate eigenvalues, while in the presence of control {\it all} the eigenvalues are stable.

\begin{figure*}
\hspace{1cm}\includegraphics[width=14cm]{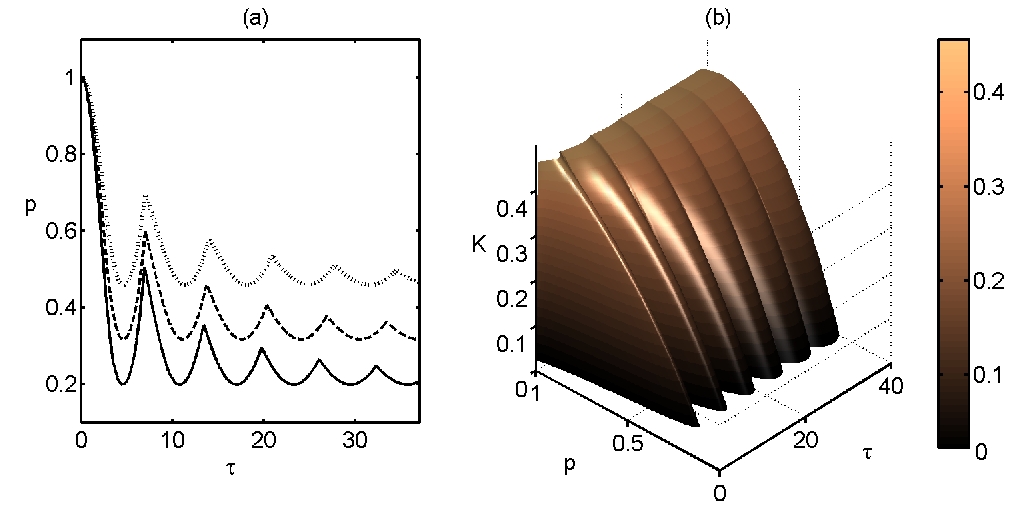}
\caption{(Color online) (a) Stability
boundary of the controlled system for $\tau_{1}=\tau_{2}=\tau$, $\zeta=0.1$ and different control strengths: $K=0$ (solid), $K=0.1$ (dashed) and
$K=0.2$ (dotted). The steady state is stable below the curves. (b) Minimum stabilizing control strength $K$ as a
function of $p$ and $\tau_{1}=\tau_{2}=\tau$.} \label{splot}
\end{figure*}

First we consider the case when the time delay of the control coincides with the time delay of the original system $\tau_{2}=\tau_{1}=\tau$. Let us determine the stability boundary Re($\lambda)=0$. Looking for solutions of Eq.~(\ref{char_eq_control}) in the form $\lambda=i\omega$ and separating real and imaginary parts, we find
\begin{equation}\label{rim_cont}
\begin{array}{l}
-\omega^{2}+1-K=\left(p\omega^{2}-K\right)\cos\omega\tau,\\
2\zeta\omega=-\left(p\omega^{2}-K\right)\sin\omega\tau.
\end{array}
\end{equation}
Squaring and adding the two equations in the last system gives
\[
\left(-\omega^{2}+1-K\right)^{2}+4\zeta^{2}\omega^{2}=\left(p\omega^{2}-K\right)^{2}.
\]
The last equation can be solved for $p$ as follows:
\[
p=\frac{1}{\omega^{2}}\left(\sqrt{\left(-\omega^{2}+1-K\right)^{2}+4\zeta^{2}\omega^{2}}+K\right).
\]
\begin{figure*}
\hspace{1cm}\includegraphics[width=14cm]{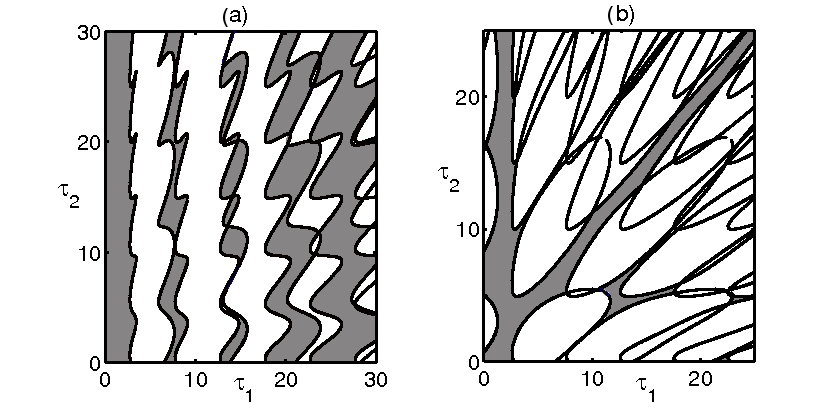}
\caption{Stability boundaries of the characteristic equation (\ref{char_eq_control}) for $p=0.4$ and $\zeta=0.1$. (a) $K=0.1$, (b) $K=0.3$. The steady state is stable inside the shaded area.}
\label{tau12P04}
\end{figure*}
Also, by dividing the second equation of the system (\ref{rim_cont}) by the first, the critical time delay can be expressed as
\[
\tau=\frac{1}{\omega}\left[\Arctan\frac{2\zeta\omega}{\omega^{2}-1+K}\pm\pi k\right],\hspace{0.5cm}
k=0,1,2,...
\]
Now, for each fixed value of the control strength $K$ and the damping $\zeta$, we have a stability boundary in the $(\tau,p)$ plane parametrized by a Hopf frequency $\omega$. Figure~\ref{splot}(a) shows that by increasing the control strength $K$, it is possible to raise the stability boundary, and thus improve overall stability of the system. This also means that those points which were unstable in the uncontrolled system, are now stable as they are moved by the control action to the area below the new stability boundary. It is worth mentioning that if $K<0$, this actually lowers the stability boundary, and hence reduces the stability.

So far, we fixed the control strength $K$ and studied the stability boundary as a curve in the $(\tau,p)$ plane. At the same time, one can consider $p$ and $\tau$ fixed (this corresponds to a single unstable fixed point of the uncontrolled system), and then solve the system (\ref{rim_cont}) for a tuple $(K,\omega)$. This would give a full picture of controllability of the system by providing the value of the minimum control strength $K$ required to stabilize the steady state for the given values of $p$ and $\tau$. The results of this computation are shown in Fig.~\ref{splot}(b), where the steady state is stable above the surface. This plot suggests that the more unstable is a steady state (i.e. the higher it is above its stability boundary in $(\tau,p)$ plane), the higher should be the control strength $K$ required to stabilize this steady state.

When $\tau_{2}\neq\tau_{1}$, the characteristic equation (\ref{char_eq_control}) has two time delays simultaneously present, which significantly complicates a stability analysis. Several approaches have been recently put forward
to study stability of equations with multiple time delays \cite{ahl04,ahl05}. Beretta and Kuang \cite{bk} have developed geometric stability switch criteria for equations with delay-dependent parameters, and this method can also be used to analyze systems with two time delays. Sipahi and Olgac \cite{SO} have suggested a systematic procedure of finding eigenvalues of characteristic equations for systems with multiple time delays by means of a substitution, which replaces an original transcendental characteristic equation with a polynomial. For the purposes of this paper, it is convenient to use a parametrization of the stability boundary in the parameter plane of two time delays, as outlined in Gu {\it et al.} \cite{Gu}. The idea is to rewrite the characteristic equation (\ref{char_eq_control}) in an equivalent form
\begin{equation}
1+a_{1}(\lambda)e^{-\lambda\tau_{1}}+a_{2}e^{-\lambda\tau_{2}}=0,
\end{equation}
where
\[
a_{1}(\lambda)=\frac{p\lambda^{2}}{1-K+2\zeta\lambda+\lambda^{2}},\hspace{0.5cm}
a_{2}(\lambda)=\frac{K}{1-K+2\zeta\lambda+\lambda^{2}}.
\]
The stability border can now be parameterized by a Hopf frequency $\omega\in\Omega$, where $\Omega=\bigcup_{i=1}^{k}\Omega_{k}$ consists of a finite number of intervals of finite length. The critical time delays at the stability boundary in the $(\tau_{1},\tau_{2})$ plane are then given by
\begin{equation}\label{tau12_forms}
\begin{array}{l}
\displaystyle{\tau_{1}=\frac{\Arg[a_{1}(i\omega)]+(2u-1)\pi\pm\theta_{1}}{\omega},}\\\\
\displaystyle{u=u_{0}^{\pm},u_{0}^{\pm}+1,u_{0}^{\pm}+2,...}\\\\
\displaystyle{\tau_{2}=\frac{\Arg[a_{2}(i\omega)]+(2v-1)\pi\pm\theta_{1}}{\omega},}\\\\
\displaystyle{v=v_{0}^{\pm},v_{0}^{\pm}+1,v_{0}^{\pm}+2,...}
\end{array}
\end{equation}
\begin{figure*}
\includegraphics[width=16cm]{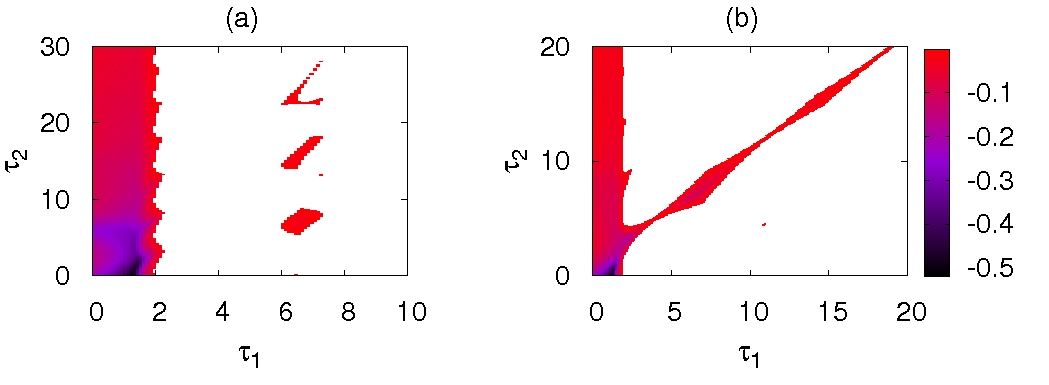}
\caption{(Color online) Real part of the leading eigenvalue of the characteristic equation (\ref{char_eq_control}) versus delay times $\tau_{1}$ and $\tau_{2}$ for
$p=0.4$ and $\zeta=0.1$. (a) $K=0.1$. (b) $K=0.3$. The color code denotes the value of Re$(\lambda)$;  only those parts are shown where it is negative. The colored areas in both plots correspond to the regions where the steady state is stable.}
\label{tau12P04eig}
\end{figure*}
where $\theta_{1},\theta_{2}\in[0,\pi]$ are calculated as
\[
\begin{array}{l}
\displaystyle{\theta_{1}=\arccos\left(\frac{1+|a_{1}(i\omega)|^{2}-|a_{2}(i\omega)|^{2}}
{2|a_{1}(i\omega)|}\right)},\\\\
\displaystyle{\theta_{2}=\arccos\left(\frac{1+|a_{2}(i\omega)|^{2}-|a_{1}(i\omega)|^{2}}
{2|a_{2}(i\omega)|}\right)},
\end{array}
\]
and $u_{0}^{\pm}$ and $v_{0}^{\pm}$ are determined as the smallest positive integers such that the corresponding $\tau_{1}^{u_{0}^{+}}$, $\tau_{1}^{u_{0}^{-}}$, $\tau_{2}^{v_{0}^{+}}$, $\tau_{2}^{v_{0}^{-}}$ are all non-negative.

\begin{figure*}
\includegraphics[width=16cm]{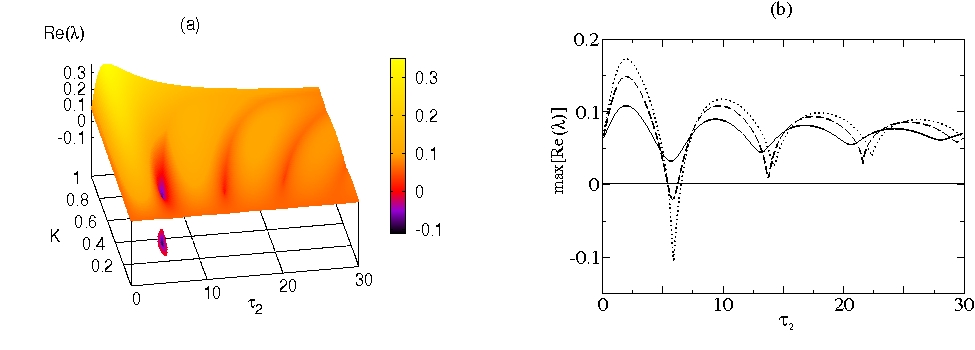}
\caption{(Color online) (a) Real part of the leading eigenvalue of the characteristic equation (\ref{char_eq_control}) for $p=0.4$, $\zeta=0.1$, and $\tau_{1}=5$.
b) Dependence of the real part of the leading eigenvalue on the control time delay $\tau_{2}$ for $K=0.15$ (solid),
$K=0.28$ (dashed), and $K=0.36$ (dotted).}
\label{KtauP04}      
\end{figure*}

Figure~\ref{tau12P04} shows the stability boundaries of the controlled system in the parameter space of the
time delays $\tau_{1}$ and $\tau_{2}$, as given by the formulas (\ref{tau12_forms}). We note that when $K=0.1$, the stability boundaries are represented
by zig-zag curves with vertical axes. For larger values of the control strength $K$,
the stability boundaries become closed curves. The steady state is stable inside the shaded areas.

The control strength $K$ appears as a parameter in $a_{1}$ and $a_{2}$, and thus determines the values
of $\tau_{1}$ and $\tau_{2}$ at the stability threshold. At the same time, in order to determine the
controllability of the system it is more convenient to fix particular values of the parameters  $\tau_{1}$,
$p$, and $\zeta$ of the original system,
and then study the stability border in the $(\tau_{2},K)$ space.
Figure~\ref{KtauP04}(a) shows how the real part of the leading characteristic eigenvalue depends
on $K$ and $\tau_{2}$, and in Fig.~\ref{KtauP04}(b) for illustrative purposes we present three possible regimes for different values
of the control strength $K$. One can observe that while for small $K$ the steady state is unstable, it becomes stable starting with some $K$, and the larger $K$ is, the more negative the real part of the leading eigenvalue becomes. It is clear from Fig.~\ref{KtauP04}(a) that the stability islands in $(\tau_{2},K)$ parameter plane are
finite, and hence if the control strength is very large, the steady state will be unstable again.

If the parameters are taken closer to the stability boundary of the uncontrolled system, a much smaller control strength $K$ is required to stabilize
an unstable equilibrium.

\begin{figure*}
\includegraphics[width=16cm]{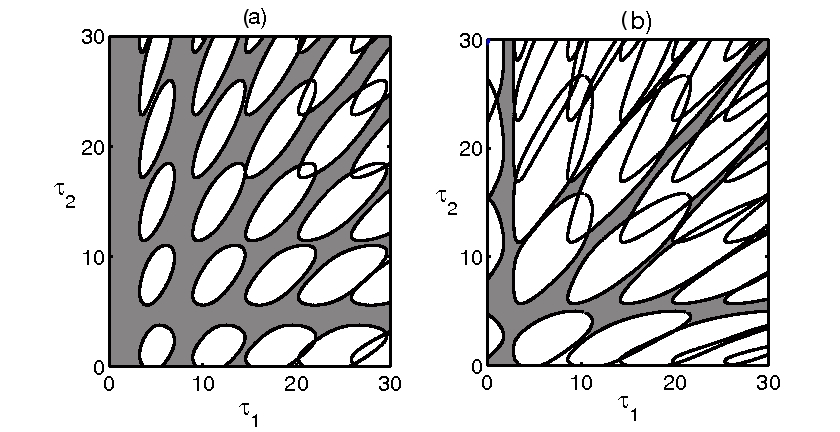}
\caption{Stability boundaries of the characteristic equation (\ref{char_eq_control}) for $p=0.25$ and $\zeta=0.1$. a) $K=0.1$, b) $K=0.3$. Shaded area corresponds to stable steady state.}
\label{tau12P025}
\end{figure*}

\begin{figure*}
\includegraphics[width=16cm]{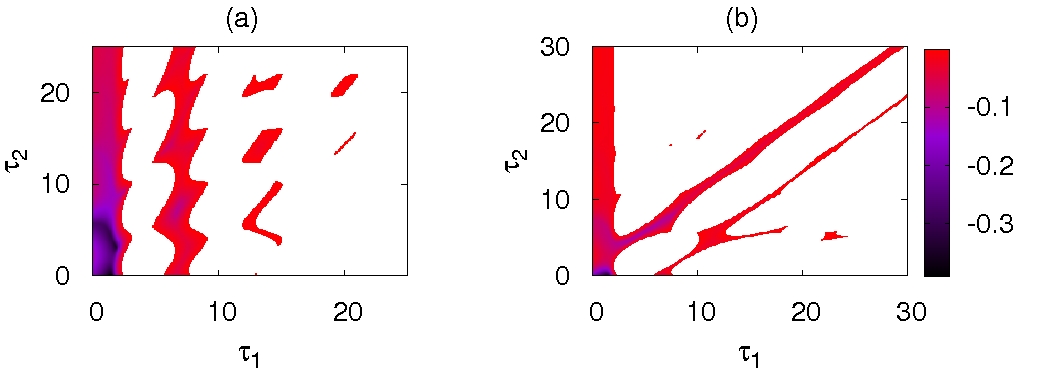}
\caption{(Color online) Real part of the leading eigenvalue of the characteristic equation (\ref{char_eq_control}) versus delay times $\tau_{1}$ and $\tau_{2}$ for $p=0.25$ and $\zeta=0.1$. (a) $K=0.1$. (b) $K=0.3$. The color code denotes the value of Re$(\lambda)$; only those parts are shown where it is negative. The colored areas in both plots correspond to the regions where the steady state is stable.}
\label{tau12P025eig}
\end{figure*}

\begin{figure*}
\includegraphics[width=16cm]{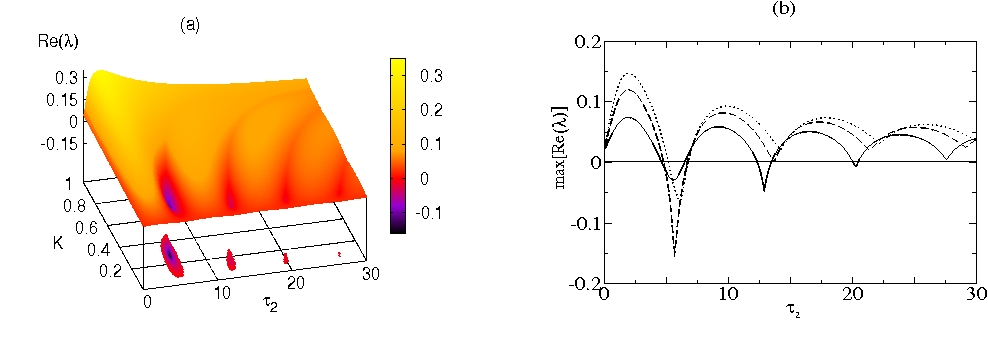}
\caption{(Color online) (a) Real part of the leading eigenvalue of the characteristic equation (\ref{char_eq_control}) for $p=0.25$, $\zeta=0.1$, and $\tau_{1}=5$.
b) Dependence of the real part of the leading eigenvalue on the control time delay $\tau_{2}$ for $K=0.15$ (solid),
$K=0.28$ (dashed), and $K=0.36$ (dotted).}
\label{KtauP025}
\end{figure*}

To deepen the understanding of the stability changes depending on the relation between the two time delays, we
have computed the maximum of the real part of the characteristic eigenvalues, which is shown in Fig.~\ref{tau12P04eig} wherever it is negative. To compute the real part of the leading eigenvalue of
the characteristic equation (\ref{char_eq_control}) we have used pseudospectral differentiation methods \cite{er,breda,bmv}. Note some of the very small stable domains (with Re($\lambda$) very close to zero) shown in Fig.~\ref{tau12P04} are not fully resolved numerically in Fig.~\ref{tau12P04eig}.
In this case all stability boundaries are closed curves shown in Fig.~\ref{tau12P025} for fixed $K=0.1$ and $K=0.3$, and the steady state is stable in the shaded area outside those curves. The corresponding plot of the real part of the leading eigenvalue, shown in Fig.~\ref{tau12P025eig}, reveals the existence of several regions of  values of $\tau_{2}$, for which
the steady state can be stabilized for the same value of the control strength $K$ and the original time delay $\tau_{1}$. If we fix $\tau_{1}$ and consider the plane of control
parameters $K$ and $\tau_{2}$, the dependence of the leading eigenvalue of the characteristic equation on these parameters is
qualitatively the same as the one for $p=0.4$, whereas the stability regions are more pronounced, see Fig.~\ref{KtauP025}. In comparison to the case of $p=0.4$, there is more than one interval of time delay $\tau_{2}$, for which stabilization of the steady state is achieved for a fixed value of the control strength $K$. Note that similar behavior as in Figs.~\ref{tau12P025} and \ref{tau12P025eig} was found for an electronic oscillator where two time delays were added in the form of two Pyragas-type time-delayed feedback control terms \cite{ahl05}.

\section{Conclusions}

For many physical, biological and engineering problems an important practical challenge is the control of stability of the steady
states. While several techniques are readily available, not all of them are equally robust and can be implemented
experimentally. In this respect, time-delayed feedback control has been demonstrated to be an efficient tool for stabilization of unstable periodic orbits and unstable steady states.

The instability of the steady states occurs quite often in time-delayed systems, where the time delay can
destabilize the steady state through a Hopf bifurcation. In this paper, we have shown how the time-delay feedback approach can be used to
stabilize unstable steady states in a neutral delay differential equation. For a given value of the control strength, it is possible to
find an analytical parametrization of (in)stability boundaries in terms of two time delays: $\tau_{1}$ (time delay of the original system)
and $\tau_{2}$ (time delay of the control force). Quite naturally, the more unstable is a steady state, as determined by the real
part of the leading eigenvalue of the corresponding characteristic equation, the higher should be the control strength required to stabilize it for a
fixed value of the control time delay $\tau_{2}$. At the same time, if the steady state is close to a Hopf bifurcation, it is easier to control it. In the latter case there is a larger number of islands in a parameter space where stabilization is successfully achieved.

It is noteworthy that the time-delayed feedback control scheme fails to achieve controllability of the steady state if $|p|>1$ independently on the intrinsic time delay and control parameters. This further highlights the difference between stabilizing unstable steady states in neutral and non-neutral time-delayed systems.

\section*{Acknowledgements}

Y.K. was supported by an EPSRC Postdoctoral Fellowship (Grant EP/E045073/1). This work was partially 
supported by Deutsche Forschungsgemeinschaft in the framework of Sfb 555. K.B. and Y.K. would like to acknowledge the hospitality of the Institut f\"ur Theoretische Physik, TU Berlin, where part of this project was completed.

\end{document}